\documentclass[prd,aps,floats,onecolumn,twoside,preprintnumbers,
superscriptaddress,floatfix,nofootinbib,showpacs]{revtex4}
\usepackage{graphicx,pstricks,rotating,amsmath,epstopdf}
\usepackage{amssymb,bbm}
\usepackage{slashed}
\begin{document}

\renewcommand{\Im}{\mathop{\rm Im}}
\setcounter{tocdepth}{3}

\title{\center{UV-IR Mixing in nonassociative Snyder $\phi^4$ theory}}

\author{Stjepan Meljanac}
\email{meljanac@irb.hr}
\affiliation{Ruder Boskovic Institute, Division of Theoretical Physics, P.O.Box 180, HR-10002 Zagreb, Croatia}
\email{meljanac@irb.hr}
\author{Salvatore Mignemi}
\affiliation{Dipartimento di Matematica e Informatica, Universita di Cagliari,\\
viale Merello 92, 091123 Cagliari, Italy}
\email{smignemi@unica.it}
\affiliation{INFN, Sezione di Cagliari, 09042 Monserrato, Italy}
\author{Josip Trampetic}
\affiliation{Ruder Boskovic Institute,  Division of Experimental Physics, P.O.Box 180, HR-10002 Zagreb, Croatia}
\email{josip.trampetic@irb.hr}
\affiliation{Max-Planck-Institut f\"ur Physik, (Werner-Heisenberg-Institut),
\\F\"ohringer Ring 6, D-80805 M\"unchen, Germany}
\email{trampeti@mppmu.mpg.de}
\author{Jiangyang You}
\email{youjiangyang@gmail.com}
\affiliation{Ruder Boskovic Institute, Division of Physical Chemistry, P.O.Box 180, HR-10002 Zagreb, Croatia}

\date{\today}

\begin{abstract}
Using a quantization of the nonassociative and noncommutative Snyder $\phi^4$ scalar field theory in a Hermitian realization, we present in this article analytical formulas for the momentum-conserving part of the one-loop two-point function of this theory in $D$-, 4-, and 3-dimensional Euclidean spaces, which are exact with respect to the noncommutative deformation parameter $\beta$. We prove that these integrals are regularized by the Snyder deformation. These results indicate that the Snyder deformation does partially regularize the UV divergences of the undeformed theory, as it was proposed decades ago. Furthermore, it is observed that different nonassociative $\phi^4$ products can generate different momentum-conserving integrals. Finally most importantly, a logarithmic infrared divergence emerges in one of these interaction terms. We then analyze sample momentum nonconserving integral qualitatively and show that it could exhibit IR divergence too. Therefore infrared divergences should exist, in general, in the Snyder $\phi^4$ theory. We consider infrared divergences at the limit $p\to 0$ as UV/IR mixings induced by nonassociativity, since they are associated to the matching UV divergence in the zero-momentum limit and appear in specific types of nonassociative $\phi^4$ products. We also discuss the extrapolation of the Snyder deformation parameter $\beta$ to negative values as well as certain general properties of one-loop quantum corrections in Snyder $\phi^4$ theory at the zero-momentum limit.
\end{abstract}

 \pacs{11.10.Nx, 11.15.-q., 12.10.-g}
%\keywords{Noncommutative Geometry, Quantum Field Theory}

\maketitle

\section{Introduction}
Several well-known arguments indicate that at very short spacetime distances the very concept of point and localizability may no longer be adequate. That this must be described by different geometrical structures is one of the oldest motivations for the introduction of noncommutative geometry~\cite{Moyal:1949sk,Connes:1994yd,Doplicher:1994tu,Landi:1997sh,Madore:1999bi,Madore:2000aq,GraciaBondia:2001tr}. The simplest kind of noncommutative geometry is the so-called canonical one~\cite{Doplicher:1994tu,Szabo:2001kg,Douglas:2001ba,Szabo:2009tn}, that can also be derived from string theories \cite{Seiberg:1999vs}.

Construction of a field theory on a noncommutative space is usually performed by deforming the product of functions (and hence of fields) with the introduction of a noncommutative star product. The noncommutative coordinates 
$\hat{x}^\mu$ satisfy
\begin{equation}
[\hat{x}^\mu,\hat{x}^\nu]=i \Theta^{\mu\nu},
\label{commutatorxMoy}
\end{equation}
where $\hat{x}^\mu$'s are Hermitian operators, and at the right-hand side of (\ref{commutatorxMoy}) the $\Theta^{\mu\nu}$ is constrained to be a real rank two antisymmetric tensor (or, more generally, an anti-Hermitian matrix as, for example, in the Wick-Voros star product.

The simplest case $\Theta^{\mu\nu}=\theta^{\mu\nu}\sim$ constant is establishing the well-known Moyal noncommutative spacetime \cite{Moyal:1949sk} (for a review see~\cite{Szabo:2001kg,Douglas:2001ba}, and references therein): $\theta^{\mu\nu}$ does not depend on $x$ (coordinate), and scales like length$^2\sim \Lambda^{-2}_{\rm NC}$, $\Lambda_{\rm NC}$ being the scale of noncommutativity, with the dimension of energy. Field theories on Moyal space admit relatively simple perturbative quantization based on a functional method, which, on the other hand, leads to a number of unconventional properties thereafter. One subtle and remarkable finding is the ultraviolet/infrared (UV/IR) mixing~\cite{Minwalla:1999px}, where the noncommtativity turns the UV divergence of the commutative theory into a matching IR divergence. Also, in the case of timelike noncommutativity, i.e., when one of the $\mu,\nu$ indices is a timelike, the noncommutative theories, in general, do not satisfy the unitarity condition \cite{Gomis:2000zz}. However, there is a case of so-called lightlike unitarity condition, $\Theta^{0i}=-\Theta^{1i}, \; \forall i=1,2,3$ \cite{Aharony:2000gz}, which is regarded as being acceptable with respect to the general unitarity condition of quantum field theories (QFTs) on noncommutative spaces. For Moyal geometry, recently it was proven that there exists a $\theta$-exact formulation of noncommutative gauge field theory based on the Seiberg-Witten map \cite{Schupp:2002up,Schupp:2008fs,Horvat:2011bs,Horvat:2011qg,Trampetic:2015zma,Horvat:2015aca} that preserves unitarity \cite{Aharony:2000gz}, and has improved UV/IR behavior at the quantum level in its supersymmetric version \cite{Martin:2016zon,Martin:2016hji,Martin:2016saw}. Noncommutativity may also have implications on cosmology, as for example through the determination of the maximal decoupling/coupling temperatures of the right-handed neutrino species in the early universe \cite{Horvat:2017gfm,Horvat:2017aqf}. Finally, an important requirement is that the quantum theory must be formulated without expansion/approximation with respect to $\Theta$, which adds considerable difficulties especially when  the Seiberg-Witten map is turned on, while yielding most trustable answers as payback.

The exact  mathematical formulation of classical field theories on geometries, like $\kappa$-Minkowski \cite{Lukierski:1991pn,Lukierski:1992dt,Meljanac:2007xb,Meljanac:2011cs,Govindarajan:2009wt,Grosse:2005iz} and Snyder \cite{Snyder:1946qz}, with respect to deformation parameters $\kappa$ and $\beta$, respectively, is also very important. However, quantum properties of these sibling theories are not as easy to characterize as the Moyal theories~\cite{Grosse:2005iz,Meljanac:2011cs}.

In his seminal paper, Snyder \cite{Snyder:1946qz}, assuming a noncommutative structure of spacetime and hence a deformation of the Heisenberg algebra, observed that it is possible to define a discrete spacetime without breaking  the Lorentz invariance. This is in contrast with the Moyal and $\kappa$-Minkowski case, where the Lorentz invariance is either broken or deformed. It is therefore interesting to investigate  the Snyder model from the point of view of noncommutative geometry. Meanwhile, new models of noncommutative geometry have been introduced \cite{Lukierski:1991pn}, and new methods, like the formalism of Hopf algebras, have been applied to their study \cite{Majid:1996kd}.

Snyder spacetime \cite{Snyder:1946qz}, the subject of the present investigation, belongs to a class of models that have been introduced and investigated using the Hopf-algebra formalism in \cite{Maggiore:1993kv,Battisti:2010sr,Meljanac:2016gbj,Meljanac:2017ikx,Meljanac:2016jwk}. These generalizations can be studied in terms of noncommutative coordinates $\hat{x}_\mu$ and momentum generators  $p_\mu$, that span a deformed Heisenberg algebra \cite{Meljanac:2017ikx}
\begin{equation}
[\hat{x}_\mu,\hat{x}_\nu]=i\beta M_{\mu\nu}\Psi(\beta p^2),\quad
[p_\mu,\hat{x}_\nu]=-i\Phi_{\mu\nu}(p),\quad
[p_\mu,p_\nu]=0,
\label{commutatorxSnyd}
\end{equation}
where Lorentz generators $M_{\mu\nu}$ satisfy standard commutation relations and $\beta$ is a real deformation parameter of dimension $\rm length^2$  $(\beta \propto\ell^2_P)$, with $\ell_P$ being the Planck length.\footnote{ For one of the $\mu,\nu$ indices being a timelike  Lorentz generator, $M_{\mu\nu}$ become simple  boost operators. However for the specific lightlike noncommutativity defined as $\Theta^{0i}=-\Theta^{1i}$  \cite{Aharony:2000gz} we get that the boost operators $M_{0i}=-M_{1i} \equiv-x_1p_i+x_ip_1, \; \forall i=1,2,3$, become pure spacelike operators of the type $\vec x\times\vec p$,  where explicit time dependence disappears and the unitarity condition is satisfied.} Functions $\Psi(\beta p^2)$ and $\Phi_{\mu\nu}(p)$ are constrained so that the Jacobi identities hold. Detailed computations and discussions of the Snyder realizations are given in previous works \cite{Battisti:2010sr,Meljanac:2017ikx}. The Snyder model has also been treated from different points of view in \cite{Girelli:2010wi,Lu:2011it,Lu:2011fh,Mignemi:2013aua,Mignemi:2015fva}. Most recently, in \cite{Meljanac:2017ikx} the construction of QFT on Snyder spacetime has finally started and some general formulation have been 
proposed, but limited to the perturbative expansion with respect to deformation parameter $\beta$ only \cite{Meljanac:2017grw}. 

A few general comments are in order. First, from the underlying mathematics, like $L_{\infty}$ algebras (see \cite{Hohm:2017cey} and references within \cite{Hohm:2017pnh}), new structures arise, for example the star-product algebra of functions, which were studied through nongeometric strings, probing noncommutative and nonassociative deformations of closed string background geometries \cite{Blumenhagen:2010hj,Lust:2010iy,Gunaydin:2016axc}, see also celebrated paper by Kontsevich \cite{Kontsevich:1997vb}. Second, the quantization of these backgrounds through explicit constructions of phase space star products were provided in \cite{Kupriyanov:2015dda,Mylonas:2012pg}, and subsequently applied to construct nonassociative theories \cite{Kupriyanov:2017oob}.  In this article we show the active role originating from the nonassociativity of the star product. The impact of these nonassociative structures on the correlation functions is also expected, so that their physical significance will be clearly visible.

In this paper, we construct the $\beta$-deformation exact Snyder $\phi^4$ action, based on the $\beta$-exact star product. This should give the same results as a summation over all orders in a perturbative expansion in $\beta$, like the one of \cite{Meljanac:2017grw}. We expect  nonperturbative quantum effects like the celebrated UV/IR mixing in Moyal space \cite{Minwalla:1999px} to reappear in this approach. Thus, the main purpose of this article is to see whether for the $\beta$-exact Snyder $\phi^4$ action these effects really occur. UV/IR mixing is, in principle, a very important quantum property and among other things, connects the noncommutative field theories with holography in a model-independent way \cite{Cohen:1998zx,Horvat:2010km}. In the literature, both holography and UV/IR mixing are known as possible windows to quantum gravity \cite{Cohen:1998zx}. In addition, recently, by using results from  \cite{Palti:2017elp}, the very notion of UV/IR mixing was interconnected with the weak gravity conjecture with scalar fields in the Lust and Palti article \cite{Lust:2017wrl}; it manifests itself as a form of hierarchical UV/IR mixing and is tied to the interaction between the weak gravity conjecture and nonlocal (possibly noncommutative) gauge operators \cite{Lust:2017wrl}.

The paper is organized as follows: in Sec. II we introduce the Hermitian realization of the Snyder algebra, the star product corresponding to this realization and the Snyder-deformed  $\phi^4$ action based on that star product. The one-loop two-point function evaluation is given in Sec. III. Next, in Sec. IV we present our arguments regarding the existence of the UV/IR mixing in Snyder $\phi^4$ theory. We discuss the effect of negative $\beta$ value on the two-point function and general properties of the one-loop quantum corrections in Snyder $\phi^4$ theory at zero-momentum limit in Sec. V. Finally, conclusions are given in Sec. VI.

\section{Exact $\phi^4$ Scalar theory in the Hermitian realization of the Snyder space}

Considering a simplified version of Eq. (\ref{commutatorxSnyd}) we write the following deformed Heisenberg algebra associated with the Snyder model as
\begin{equation}
[\hat{x}_\mu,\hat{x}_\nu]=i\beta M_{\mu\nu},\quad
[p_\mu,\hat{x}_\nu]=-i(\eta_{\mu\nu}+\beta p_\mu p_\nu),\quad
[p_\mu,p_\nu]=0.
\label{2.1}
\end{equation}

The  star product for the Hermitian Snyder realization is given by~\cite{Meljanac:2016gbj,Meljanac:2017ikx}
\begin{equation}
e^{ikx}\star e^{iqx}=e^{iD(k,q)x}e^{iG(k,q)},
\label{starproduct}
\end{equation}
with the following exact expressions for $D_\mu(k,q)$ and $G(k,q)$:
\begin{gather}
D_\mu(k,q)=\frac{1}{1-\beta k\cdot q}\Bigg(\bigg(1-\frac{\beta k\cdot q}{1+\sqrt{1+\beta k^2}}\bigg)k_\mu+\sqrt{1+\beta k^2}q_\mu\Bigg),
\label{snyderrealization}
\\
G(k,q)=i\frac{D+1}{2}\ln\left(1-\beta k\cdot q\right),
\label{6}
\end{gather}
where $D$ is dimension of the spacetime.

The Snyder momentum addition $D_\mu(k,q)$ satisfies the following relations
\begin{equation}
D_\mu(k,-k)=0,\;D_\mu(k,0)=k,\;D_\mu(0,q)=q.
\label{relations}
\end{equation}

Taking into account the integration by part identity, there are three possible candidates for the Snyder-exact $\phi^4$ interaction:
\begin{equation}
S_1=\frac{\lambda}{4!}\int \phi\big(\phi\star(\phi\star\phi)\big),
\label{S_1}
\end{equation}
\begin{equation}
S_2= \frac{\lambda}{4!}\int \phi\big((\phi\star\phi)\star\phi\big),
\label{S_2}
\end{equation}
\begin{equation}
S_3=\frac{\lambda}{4!}\int (\phi\star\phi)(\phi\star\phi).
\label{S_3}
\end{equation}
The arbitrary linear combination of these three terms can be taken to write the general Snyder-exact $\phi^4$ interaction:
\begin{equation}
S=c_1 S_1 + c_2 S_2 + c_3 S_3
,\; c_1+c_2+c_3=1.
\label{S}
\end{equation}
Normalization $c_1+c_2+c_3=1$ is introduced to recover the conventional $\phi^4$ interaction in the commutative limit.

\begin{figure}[t]
\begin{center}
\includegraphics[width=4cm,angle=0]{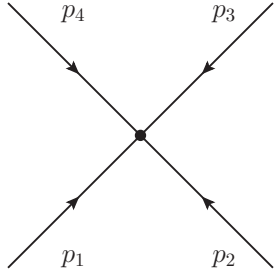}
\end{center}
\caption{Four-point Feynman rule.}
\label{fig:FR}
\end{figure}

\section{ONE-LOOP TWO-POINT FUNCTIONS}

\begin{figure}[t]
\begin{center}
\includegraphics[width=6cm,angle=0]{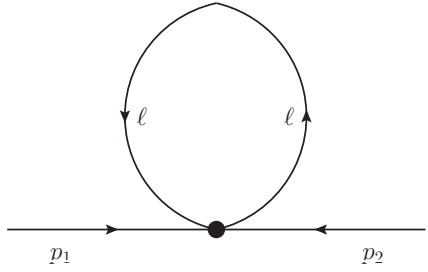}
\end{center}
\caption{The one-loop tadpole contribution to the two-point function.}
\label{fig:FD1}
\end{figure}

Using the functional method in momentum space, the generating functional of the scalar field theory with $\phi^4$ interactions on the Snyder deformed Euclidean space can be defined~\cite{Meljanac:2017grw}. Considering Figs. \ref{fig:FR} and \ref{fig:FD1}, the one-loop two-point function is then given by
\begin{equation}
\begin{split}
G_{1-loop}(x_1,x_2)=&-\frac{1}{2}\frac{\lambda}{4!}\int\frac{d^D p_1}{(2\pi)^D}\frac{d^D p_2}{(2\pi)^D}\frac{d^D \ell}{(2\pi)^D}
\frac{e^{ip_1x_1}}{p_1^2+m^2}\frac{e^{ip_2x_2}}{p_2^2+m^2}\frac{(2\pi)^D}{\ell^2+m^2}
\\&\cdot\sum\limits_{\sigma\in P_4}\delta\Big(D_4\big(\sigma\big(p_1,p_2,\ell,-\ell\big)\big)\Big)\cdot g_3\Big(\sigma\big(p_1,p_2,\ell,-\ell\big)\Big).
\end{split}
\label{1l2p}
\end{equation}
The general definitions of $\delta(D_4(\sigma(p_1,p_2,\ell,-\ell)))$ and $g_3(\sigma(p_1,p_2,\ell,-\ell))$ are given in \cite{Meljanac:2017grw}. In particular, $\sigma$ denotes a permutation $P_4$ over the momenta in both
arguments, respectively. We are going to compute some of them for each of the interactions $S_1,S_2$, and $S_3 $ in the following subsections.

\subsection{Momentum-conserving integrals}

\subsubsection{The one-loop two-point function from the action $S_1$}

Out of 24 $P_4$ permutations, we observe that the equation $D_4=0$ for the arguments $(p_1,p_2,\ell,-\ell)$, $(p_1,p_2,-\ell,\ell)$, $(p_2,p_1,\ell,-\ell)$, $(p_2,p_1,-\ell,\ell)$, $(\ell,-\ell,p_1,p_2)$,
$(-\ell,\ell,p_1,p_2)$, $(\ell,-\ell,p_2,p_1)$ and $(-\ell,\ell,p_2,p_1)$ admit the same unique solution $p_1=-p_2$. The remaining sixteen can be shown to be momentum nonconserving by checking the iterative solution to the
$\delta$ functions up to the $\beta^2$ order. We then work on the Jacobian determinant with respect to $p_2$ and the $g_3$ factor. For the first four cases we have
\begin{equation}
\det\left(\frac{\partial D_{4_\mu}}{\partial p_{2_\nu}}\right)\bigg|_{p_2=-p_1}=1,\quad g_3=\left(1+\beta\ell^2\right)^{-\frac{D+1}{2}},
\label{4.2}
\end{equation}
while, for the others,
\begin{equation}
\begin{split}
\det\left(\frac{\partial D_{4_\mu}}{\partial p_{2_\nu}}\right)\bigg|_{p_2=-p_1}=&\det\left(\frac{\partial D_\mu}{\partial q_\nu}\right)\bigg|_{\stackrel{q=0}{k=\pm \ell}}
\cdot\det\left(\frac{\partial D_\mu}{\partial k_\nu}\right)\bigg|_{\stackrel{q=-k}{k=\pm p_1}}
\\
=&\det\left(\frac{\partial D_\mu}{\partial q_\nu}\right)\bigg|_{\stackrel{q=0}{k=\pm \ell}}
\cdot
\det\left(\frac{\partial D_\mu}{\partial q_\nu}\right)\bigg|_{\stackrel{q=-k}{k=\pm p_1}}
\\
=&\left(1+\beta\ell^2\right)^{\frac{D+1}{2}}\left(1+\beta p_1^2\right)^{-\frac{D+1}{2}},
\\
g_3=&\left(1+\beta p_1^2\right)^{-\frac{D+1}{2}}.
\end{split}
\label{4.3}
\end{equation}
Thus all eight integrals take the same form and sum to
\begin{equation}
\begin{split}
&-\frac{\lambda}{6}\int\frac{d^D p_1}{(2\pi)^D}\frac{d^D p_2}{(2\pi)^D}\frac{d^D \ell}{(2\pi)^D}\cdot
\frac{e^{ip_1x_1}}{p_1^2+m^2}\frac{e^{ip_2x_2}}{p_2^2+m^2}\frac{1}{\left(\ell^2+m^2\right)\left(1+\beta\ell^2\right)^{\frac{D+1}{2}}}.
\end{split}
\label{4.4}
\end{equation}

We then extract a universal tadpole integral for the momentum-conserving part of the Snyder two-point function
\begin{equation}
\begin{split}
\mathcal I_1=\int\frac{d^D\ell}{(2\pi)^D}\frac{1}{\left(\ell^2+m^2\right)\left(1+\beta\ell^2\right)^{\frac{D+1}{2}}}.
\end{split}
\label{I1}
\end{equation}
One can immediately notice that this integral is UV finite in any dimension because of the fast $\mathcal O\left(|\ell|^{-(D+1)}\right)$ damping term $(1+\beta\ell^2)^{-\frac{D+1}{2}}$. For general $D$ the integral can be expressed analytically using hypergeometric functions
\begin{equation}
\mathcal I_1=(4\pi)^{-\frac{D}{2}}\beta^{1-\frac{D}{2}}\left(\frac{\sqrt{\pi}}{\Gamma\left(\frac{D+1}{2}\right)}\frac{1}{m^2\beta}\;{_2F_1}\left(1,\frac{D}{2};\frac{1}{2};\frac{1}{m^2\beta}\right)-\frac{(m^2\beta)^{\frac{D}{2}-1}}{(m^2\beta-1)^{\frac{D+1}{2}}}\frac{\pi}{\Gamma\left(\frac{D}{2}\right)}\right).
\label{4.6}
\end{equation}
Further simplifications occur for specific values of $D$. When $D=4$,
\begin{equation}
\mathcal I_1\Big|_{D=4}=\frac{1}{(4\pi)^2\beta}\frac{2}{3}\left(\frac{1+2m^2\beta}{(1-m^2\beta)^2}-\frac{3m^2\beta}
{\sqrt{(m^2\beta-1)^5}}\sec^{-1}\sqrt{m^2\beta}\right),
\label{I14}
\end{equation}
and when $D=3$
\begin{equation}
\mathcal I_1\Big|_{D=3}=\frac{1}{8\pi\sqrt\beta}\frac{1}{(1+ m\sqrt\beta)^2}.
\label{I13}
\end{equation}

\subsubsection{The one-loop two-point function from the action $S_2$}

We now move from $S_1$ to $S_2$. There are again eight momentum-conserving permutations for $S_2$: they are $(p_1,\ell,-\ell,p_2)$, $(p_1,-\ell,\ell,p_2)$, $(p_2,\ell,-\ell,p_1)$, $(p_2,-\ell,\ell,p_1)$, $(\ell,p_1,p_2,-\ell)$, $(-\ell,p_1,p_2,\ell)$, $(\ell,p_2,p_1,-\ell)$ and $(-\ell,p_2,p_1,\ell)$. The first four of them have Jacobian equal to one and $g_3=(1+\beta\ell^2)^{-\frac{D+1}{2}}$, while the others satisfy
\begin{equation}
\begin{split}
\det\left(\frac{\partial D_{4_\mu}}{\partial p_{2_\nu}}\right)\bigg|_{p_2=-p_1}=&\det\left(\frac{\partial D_\mu}{\partial k_\nu}\right)\bigg|_{\stackrel{k=0}{q=\pm\ell}}
\cdot
\det\left(\frac{\partial D_\mu}{\partial k_\nu}\right)\bigg|_{\stackrel{k=-q}{q=\mp p_1}}
\\=&\det\left(\frac{\partial D_\mu}{\partial k_\nu}\right)\bigg|_{\stackrel{k=0}{q=\pm\ell}}
\cdot
\det\left(\frac{\partial D_\mu}{\partial q_\nu}\right)\bigg|_{\stackrel{k=-q}{q=\mp p_1}}
\\
=&\left(1+\beta\ell^2\right)\left(1+\beta p_1^2\right)^{-\frac{D+1}{2}},
\\
g_3=&(1+\beta p_1^2)^{-\frac{D+1}{2}}.
\end{split}
\label{4.9}
\end{equation}
We then have a second type of tadpole integral
\begin{equation}
\mathcal I_2=\int\frac{d^D\ell}{(2\pi)^D}\frac{1}{(\ell^2+m^2)(1+\beta\ell^2)},
\label{I2}
\end{equation}
which converges only when $D<4$. This integral can nevertheless be evaluated using the standard dimensional regularization prescription
\begin{equation}
\begin{split}
\mathcal I_2=&\int\frac{d^D\ell}{(2\pi)^D}\frac{1}{(\ell^2+m^2)(1+\beta\ell^2)}
=\beta^{1-\frac{D}{2}}\int\frac{d^D\ell}{(2\pi)^D}\frac{1}{(\ell^2+\beta m^2)(1+\ell^2)}
\\=&\beta^{1-\frac{D}{2}}\frac{\Gamma\left(2-\frac{D}{2}\right)}{(4\pi)^{\frac{D}{2}}}\int\limits_0^1 dx\,\Big(1-x\big(1-\beta m^2\big)\Big)^{\frac{D}{2}-2}
=-\beta^{1-\frac{D}{2}}\;\frac{\Gamma\left(1-\frac{D}{2}\right)}{(4\pi)^{\frac{D}{2}}}\;\frac{1-(\beta m^2)^{\frac{D}{2}-1}}{1-\beta m^2}.
\end{split}
\label{I2D}
\end{equation}
In the $D\to 4-\epsilon$ limit this integral reduces to
\begin{eqnarray}
\mathcal I_2\Big|_{D\to 4-\epsilon}&=&\frac{1}{(4\pi)^2\beta}\left(\frac{2}{\epsilon}+1-\gamma_E+\ln(4\pi\beta)+\frac{m^2\beta}{1-m^2\beta}\ln\big(m^2\beta\big) \right)+\mathcal O(\epsilon).
\label{I24}
\end{eqnarray}
On the other hand, when $D=3$, $\mathcal I_2$ takes a simple finite value
\begin{equation}
\mathcal I_2\Big|_{D=3}=\frac{1}{4\pi\sqrt\beta}\frac{1}{1+m\sqrt\beta},
\label{I23}
\end{equation}
which is regularization independent.

\subsubsection{The one-loop two-point function from the action $S_3$}

An even more complicated situation occurs with the third term of the interaction (\ref{S}). There we have 16 different momentum-conserving permutations. Twelve of them take the same form as $\mathcal I_1$. There are four other momentum-conserving permutations $(p_1,\ell,p_2,-\ell)$, $(p_1,-\ell,p_2,\ell)$, $(p_2,\ell,p_1,-\ell)$ and $(p_2,\ell,p_1,-\ell)$ which are different, because the determinant $\det\left(\frac{\partial D_\mu}{\partial k_\nu}\right)$ must be evaluated at a general point $\left(k=\pm p, q=\pm \ell\right)$ with $p$ being an external momentum, as shown in Appendix A. The result leads to the following loop integral
\begin{equation}
\begin{split}
\mathcal I_3=&\int\frac{d^D\ell}{(2\pi)^D}\frac{1}{(\ell^2+m^2)(1+\beta\ell^2)}\left(1-\frac{\beta p\cdot \ell}{1+\sqrt{1+\beta p^2}}\right)^{1-D}.
\end{split}
\label{I3D}
\end{equation}
This integral can be evaluated using dimensional regularization techniques, as demonstrated in  Appendix B. The result clearly shows that there is no $1/\epsilon$ UV divergence in the $D\to 4$ limit; however, a logarithmic IR divergent term $\ln\big(\beta p^2\big)$ does emerge in the same limit. Since taking away the last, momentum dependent factor from \eqref{I3D} simply turns $\mathcal I_3$ back to $\mathcal I_2$, we conclude that an effect of this factor is to turn the UV divergence in $\mathcal I_2$ into an IR divergence, or, in other words, to induce UV/IR mixing.

\subsection{Momentum nonconserving integrals}

\subsubsection{General considerations of the momentum nonconserving integral}

In previous subsections III.A1, III.A2, and III.A3,  we have only considered 
the momentum-conserving integrals $\mathcal I_1$, $\mathcal I_2$, and $\mathcal I_3$, respectively. There are, as already discussed in~\cite{Meljanac:2017grw} and in the prior 
parts of this article, a number of momentum nonconserving ones as well. Unlike the momentum-conserving integrals, we do not have explicit integrated expression for such integrals. Nevertheless, we shall presently discuss certain of their properties which may be accessible without full integration.

Before we start our technical discussion it is also worthy to mentioning that momentum nonconservation, causing loss of translation invariance, could be a much more fundamental issue regarding certain basis of quantum field theory, which we are not going to study in this article. Instead we follow the prescription in~\cite{Meljanac:2017grw} to eliminate the deformed $\delta$ functions in \eqref{1l2p} by an integration over one fixed external momentum, here $p_2$, i.e., making it a function of $(p_1,\ell)$: $p_2^\sigma(p_1,\ell)$ with $\sigma$ permutations. The to-be-evaluated loop integral would formally bear the following form:
\begin{equation}
\begin{split}
G_{1-loop}(x_1,x_2)=&-\frac{1}{2}\frac{\lambda}{4!}\int\frac{d^D p_1}{(2\pi)^D}\frac{d^D \ell}{(2\pi)^D}
\\&\cdot
\frac{e^{ip_1x_1}}{p_1^2+m^2}\frac{1}{\ell^2+m^2}
%\\&
\cdot\sum\limits_{\sigma\in P_4}\frac{e^{ip_2^\sigma(p_1,\ell)x_2}}{p_2^\sigma(p_1,\ell)^2+m^2} g_3\Big(\sigma\big(p_1,p_2^\sigma(p_1,\ell),\ell,-\ell\big)\Big)\left(\frac{\delta D_4}{\delta p_2}\right)^{-1}.
\end{split}
\label{formal}
\end{equation}
At this point, the remaining task is to solve each of the $\sigma$-permuted  functions $p_2^\sigma(p_1,\ell)$ explicitly, which is not really easy though. What one could try is to use the fact that star product \eqref{starproduct} contains only vector objects (vectors and scalar products), therefore it could be convenient to project $\ell$ and $p_2$ momenta to the external moment $p_1$ component and to the perpendicular component $\ell_{\perp}$ of loop moment $\ell$, making it $\ell=(x p_1,\ell_{\perp})$, and set up a simple ansatz for $p_2$ as:  $p_2=f_1(x,p_1,\ell_{\perp}) p_1+f_2(x,p_1,\ell_{\perp}) \ell_{\perp}$. With this setting the Snyder-deformed momentum conservation in one-loop two-point function becomes a set of algebraic equations with respect to $f_1$ and $f_2$, respectively. 

\subsubsection{Superficial UV divergence of the momentum nonconserving integral}

As an example, we study the nonconserving integral with modified $\delta$ function $\delta(D\left(p_2,-\ell)+D(\ell,p_1)\right)$ coming from $S_3$ term (\ref{S_3}). In this case the momentum $p_2$ is not equal to $-p_1$ from Fig.\ref{fig:FD1}, and the difference starts at $\beta^1$ order. The relevant integral is given below:
\begin{equation}
I=\int\frac{d^D\ell}{(2\pi)^D}\frac{1}{\ell^2+m^2}\frac{e^{ip_2 x_2}}{p_2^2+m^2}\frac{1}{1+\beta\ell^2}\left(1+\frac{\beta \ell\cdot p_2}{1+\sqrt{1+\beta p_2^2}}\right)^{1-D}\left(\frac{1+\beta\ell\cdot p_2}{1-\beta\ell\cdot p_1}\right)^{\frac{D+1}{2}},
\label{sample}
\end{equation}
where defining an equation for $p_2$, in accordance to the ansatz setup,
\begin{equation}
p_2^{\sigma}(p_1,\ell)=f^{\sigma}_1(x,\ell_{\perp},p_1)p_1+f^{\sigma}_2(x,\ell_{\perp},p_1)\ell_{\perp},\;\forall \sigma\in P_4-{\rm permutations},
\label{28}
\end{equation}
is resolved by using the Snyder momentum addition relations (\ref{relations})
\begin{equation}
D_\mu(p_2,-\ell)=-D_\mu(\ell,p_1).
\label{29}
\end{equation}
The above simple equation (\ref{29}) then, after using (\ref{28}), transfers into two complicated algebraic relations for, generically $f_1$ and $f_2$, respectively,
\begin{gather}
\frac{1}{1+\beta(x f_1 p_1^2+f_2\ell_{\perp}^2)}\left(f_1-x+\frac{\beta\ell_{\perp}^2(f_1 f_2-f_2^2)}{1+\sqrt{1+\beta(f_1^2p_1^2+f_2^2\ell_{\perp}^2)}}\right)=-\frac{1}{1-x\beta p_1^2}\left(1+x+\frac{\beta\ell_{\perp}^2}{1+\sqrt{1+\beta(x^2p_1^2+\ell_{\perp})}}\right),
\\
\frac{1}{1+\beta(x f_1 p_1^2+f_2\ell_{\perp}^2)}\left(f_2-1+\frac{\beta p_1^2(xf_1f_2-f_1^2)}{1+\sqrt{1+\beta(f_1^2p_1^2+f_2^2\ell_{\perp}^2)}}\right)=-\frac{1}{1-x\beta p_1^2}\left(1-\frac{x\beta p_1^2}{1+\sqrt{1+\beta(x^2p_1^2+\ell_{\perp})}}\right).
\end{gather}
While it is hard to obtain closed form solution for $f_1$ and $f_2$ from these two equations, one can use them to analyze the large $|\ell|$ behavior of $f_1$ and $f_2$ by realizing that $x\sim\ell_{\perp}\sim |\ell|^1$. Using an ansatz $f_i\sim|\ell|^{k_i}$ for large $|\ell|$, we find that $k_1=1$ and $k_2=0$, i.e., $p_2\sim |\ell|^1$. Using this scaling we find that \eqref{sample} is superficially UV finite for $D<9$. Thus the integral \eqref{sample} is superficially UV finite at four dimensions.

Finally full solutions, if obtainable, are lengthy and yet the analytical solution to the integration over $\ell$ in (\ref{sample}) is still at large. We hope such integral could be solved in near future.

\section{UV/IR mixing}

Generally speaking, when a UV-divergent loop integral is regularized by deformation, turning off the deformation would lead to divergences in the commutative limit at the quantum level. For example, in the Moyal $\phi^4$ theory, the nonplanar/regularized integral of the two-point function reads
\begin{equation}
I(p,\theta)=\int\frac{d^D\ell}{(2\pi)^D}\frac{e^{-\frac{1}{2}{\ell_{\mu}\theta^{\mu\nu}p_\nu}}}{\ell^2+m^2}.
\label{61}
\end{equation}
When either $\theta$ or $p$ goes to zero in the integrand, this integral becomes the UV divergent commutative tadpole
\begin{equation}
I(p,0)=I(0,\theta)=\int\frac{d^D\ell}{(2\pi)^D}\frac{1}{\ell^2+m^2}.
\label{62}
\end{equation}
For this reason one expects that the integral $I(p,\theta)$ would exhibit a $(\theta p)$-dependent divergence, which was indeed found~\cite{Minwalla:1999px}.

A more careful look at the discussion above forces us to conclude that we are here actually considering two divergences, $\theta\to 0$ and $p\to 0$. The first divergence, occurring at the commutative limit $\theta\to 0$, is less surprising since an UV divergence is already present in the undeformed theory. So, this limit could be simply interpreted as a recovery.\footnote{The situation is different when the interaction is purely noncommutativity originated, for example in the Moyal U(1) (S)YM. There the UV divergence at the commutative limit is also an anomaly, since the undeformed theory is free.}

The second divergence $p\to 0$ is more intriguing. It shifts the UV divergence in the commutative/undeformed theory to the IR regime ($p\to 0$), which is a big modification to the quantum field theory, and leads to the (in-)famous UV/IR
mixing. It is not hard to see that the reason why these two divergences become associated with each other in the Moyal theory is the regulator  momentum dependence, which is a consequence of the tensorial nature of the Moyal deformation parameter $\theta^{\mu\nu}$.

Now we move from Moyal to the Snyder $\phi^4$ theory. First we consider the same vanishing external momentum limit of \eqref{1l2p}. Since such a limit brings \eqref{61} to \eqref{62}, it could be considered as an indicator of the UV behavior without momentum-dependent regularization.  Using \eqref{relations} it is not hard to find that, in the limit of vanishing external momenta and for any normalized combination of $c_1$, $c_2$, and $c_3$, the integrand of \eqref{1l2p} satisfies the following relation
\begin{equation}
\begin{split}
\lim\limits_{p_1\to 0}-\frac{1}{2}\frac{\lambda}{4!}&\int{(2\pi)^D}\frac{d^D p_2}{(2\pi)^D}\frac{d^D \ell}{(2\pi)^D}
\frac{e^{ip_1x_1}}{p_1^2+m^2}\frac{e^{ip_2x_2}}{p_2^2+m^2}\frac{(2\pi)^4}{\ell^2+m^2}
\\&\cdot\sum\limits_{\sigma\in P_4}\delta\Big(D_4\big(\sigma\big(p_1,p_2,\ell,-\ell\big)\big)\Big)\cdot g_3\Big(\sigma\big(p_1,p_2,\ell,-\ell\big)\Big)
\sim-\frac{\lambda}{6}\big(2\mathcal I_1+\mathcal I_2\big).
\end{split}
\label{1l2p1}
\end{equation}
In other words the $4!=24$ momentum permutations reduce to only two types: 16 $\mathcal I_1$ and 8 $\mathcal I_2$, respectively. Therefore, the zero-momentum limit is UV divergent because of the $\mathcal I_2$ integral. However, $\mathcal I_2$ is, unlike the Moyal theory, regulated from the commutative quadratic to logarithmic UV divergences. From this observation we conclude that the commutative and IR limits discussed above become independent from each other in Snyder theory, since the Snyder deformation parameter $\beta$ is a scalar. The $\beta\to 0$ limit of the integrals $\mathcal I_1$ and $\mathcal I_2$ is divergent, which is the expected recovery of the divergence of the commutative theory.

Next we turn to the momentum dependence of the Snyder one-loop two-point functions. We have only computed one part of it as the integral $\mathcal I_{3}$. When $D\ge 4$, this integral is superficially UV finite only when the external
momentum $p\ne 0$, and therefore it can exhibit an infrared divergence when $p\to 0$. Indeed, once we evaluate $\mathcal I_{3}$ properly, we find in the limit $p\to 0$, \footnote{A concern remains that the logarithm is complex in Euclidean spacetime. It is not so for timelike momenta ($p^2<0$) in the Minkowski spacetime with the (-- + + +) metric which is compatible with the deformation we use~\cite{Meljanac:2017ikx}.}
\begin{equation}
\mathcal I_{3}
\,\bigg|_{\hspace{.5mm}{\substack
{D=4\\p\to 0}}}\;
\sim \;-\frac{1}{(4\pi)^2\beta}
\ln\big(-\beta p^2\big),
\label{I3H40}
\end{equation}
and hence a $\ln p^2$ divergence when $p\to 0$. (See Appendix B for more details.)  This UV/IR mixing may be considered as a new type induced by nonassociativity in comparison with the Moyal $\phi^4$ theory, since the corresponding UV divergence is of the $\mathcal I_2$ type, i.e. already regulated by the Snyder deformation when compared with the commutative theory, while the additional momentum-dependent regulator in $\mathcal I_{3}$ \eqref{I3D} comes as a consequence of the nonassociativity of the Snyder star product.

More generally speaking, the full four-dimensional Snyder one-loop two-point function, including the momentum nonconserving integrals, depends on the external momentum. Although we do not know explicitly the results of the momentum nonconserving integrals, we do know that when taking the external momentum zero limit at the integrand level, part of them, for example \eqref{sample} discussed in Sec. III.B.2., becomes the UV divergent integral $\mathcal I_2$ (\ref{I2}) in accord with the aforementioned universal zero-momentum limit \eqref{1l2p1}. On the other hand, when $p=p_1\neq 0$, this integral exhibits a superficially finite UV power counting divergence as shown in Sec. III.B.2., so we may qualitatively conclude that (some of) those integrals in the full one-loop two-point function, which converge to $\mathcal I_2$ in the $p\to 0$ limit, would exhibit IR divergence in 4D if they are finite, at finite nonzero value of the momentum $p$. Therefore, we consider UV/IR mixing as a general property of the Snyder one-loop two-point function.

\section{Discussions}

Before concluding the article, it is worth noting that a variant of our model exists for $\beta<0$~\cite{Mignemi:2011gr}. In such a case, the momenta are bounded by $p^2<1/\beta$ and the integrals run over a finite range.
We can express the external momentum-independent integrals $\mathcal I_1$
and $\mathcal I_2$ in this case by introducing the adimensional loop momenta $L^\mu=\sqrt{|\beta|}\ell^\mu$. The integrals then run from $L=|L^\mu|=0$ to $L=1$.
It is not hard to show that  in this setting $\mathcal I_2$ reduces to the following expression
\begin{equation}
\mathcal I_2=\frac{2\beta}{\Gamma\left(\frac{D}{2}\right)}(4\pi\beta)^{-\frac{D}{2}}\int\limits_0^1 dL\, \frac{L^{D-1}}{(L^2+m^2)(1-L^2)}.
\end{equation}
The integral above is divergent at its upper limit for any $D$, and therefore it is not regularizable by dimensional regularization.
Similarly, $\mathcal I_1$  becomes
\begin{equation}
\mathcal I_1=\frac{2\beta}{\Gamma\left(\frac{D}{2}\right)}(4\pi\beta)^{-\frac{D}{2}}\int\limits_0^1 dL\, \frac{L^{D-1}}{(L^2+m^2)(1-L^2)^{\frac{D+1}{2}}}.
\end{equation}
This integral is still superficially divergent at the upper limit, and one can only assign it a dimensionally regularized value in terms of the Gauss hypergeometric function:
\begin{equation}
\mathcal I_1=\frac{\Gamma\left(\frac{1-D}{2}\right)\Gamma\left(\frac{D}{2}\right)}{2m^2\Gamma\left(\frac{1}{2}\right)}
\,_2F_1\left(1,\frac{D}{2};\frac{1}{2};\frac{-1}{m^2}\right).
\end{equation}
This expression is finite for even dimensions, but divergent for odd dimensions as $\Gamma\left(\frac{1-D}{2}\right)$. Technically, the complicated divergences we have encountered in this section are not surprising as the cutoff at $L=1$ still leaves the same pole at the upper boundary of the integral. It appears that the negative $\beta$ case could be more difficult to handle at loop level than positive $\beta$. Thus we leave this issue for future investigation.

Another technical possibility is to define and compute an analogue  
to the Coleman-Weinberg effective potential~\cite{Coleman:1973jx} by using the same zero external momentum limit integrand in Eq. \eqref{1l2p1}, \footnote{This means that here we completely ignore the UV/IR mixing issue, yet, as we will see, there is still another impact from Snyder deformation. We also consider zero mass for simplicity.} which yields
\begin{equation}
\begin{split}
V_{\rm eff}(\varphi)=\frac{\lambda}{4!}\varphi^4
+\beta^{-\frac{D}{2}}\int\frac{d^D\ell}{(2\pi)^D}\frac{1}{2}\sum\limits_{n=1}^{\infty}\frac{(-1)^{n+1}}{n}\Bigg(\frac{\beta\lambda\varphi^2}{6\ell^2(1+\ell^2)}\bigg(1+\frac{2}{(1+\ell^2)^{\frac{D-1}{2}}}\bigg)\Bigg)^n
\\=\frac{\lambda}{4!}\varphi^4+\frac{1}{2}\beta^{-\frac{D}{2}}\int\frac{d^D\ell}{(2\pi)^D}\ln\Bigg(1+\frac{\beta\lambda\varphi^2}{6\ell^2(1+\ell^2)}\bigg(1+\frac{2}{(1+\ell^2)^{\frac{D-1}{2}}}\bigg)\Bigg),
\end{split}
\label{effectivepotential}
\end{equation}
where $\varphi$ denotes the constant-valued field in  the zero-momentum limit.

\begin{figure}[t]
\begin{center}
\includegraphics[width=9cm,angle=0]{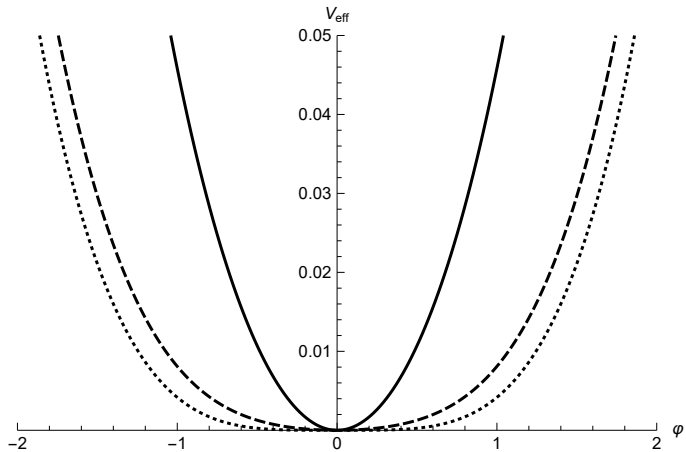}
\end{center}
\caption{The one-loop corrected effective potential of Snyder deformed three-dimensional $\phi^4$ theories for $\lambda=0.1$ and $\beta=0.001$ (solid line), $\lambda=\beta=0.1$ (dashed line) and the tree level effective potential for $\lambda=0.1$ (dotted line). The smaller $\beta$ value clearly increases the deviation from the tree level value. One may also notice that unlike the commutative theory~\cite{Peskin:1995ev}, no artificial new minima  emerge despite large loop correction.}
\label{fig:Veff}
\end{figure}

As already suggested by \eqref{I1} and \eqref{I2}, the above loop integral (\ref{effectivepotential}) is finite and computable for $D=3$, giving
\begin{equation}
V_{\rm eff}(\varphi)\Big|_{D=3}=\frac{\lambda}{4!}\varphi^4+\frac{1}{12\pi^2\sqrt{\beta^3}}\bigg(2\pi-\sum\limits_{i=1}^6f(x_i)\bigg),
\label{Veffdef}
\end{equation}
where
\begin{equation}
f(x)=\frac{2(a-1)x^4+(7a-2)x^2 -6 a}{3 x^5+4 x^3+(1+a)x}\ln (-x),\quad a=\frac{1}{6}\beta\lambda\varphi^2,
\label{6.4}
\end{equation}
and $x_{1,2,...,6}$ are the solutions to the sixth-order polynomial equation
\begin{equation}
x^6+2x^4+(1+a)x^2+3a=0.
\label{6.5}
\end{equation}
Once we analyze \eqref{Veffdef} numerically, it is not hard to find out that the aforementioned $\beta\to 0$ divergence could be considered as enhancing the one-loop contribution against the tree level at small $\beta$ values, as illustrated in Fig.~\ref{fig:Veff}.
In general, large loop corrections suggest that certain nonperturbative effects may occur~\cite{Peskin:1995ev}. One natural question is then whether it is possible to follow the Wilson-Fisher $4-\epsilon$ analysis~\cite{Wilson:1971dc,Glimm:1987ng,Kleinert:2001ax} instead of going naively to dimension three, as the Wilson-Fisher approach is long proven to be correct in extracting the critical behavior of $\phi^4$ and related theories in lower dimensions. Besides the large perturbation issue, this procedure is not yet possible since we do not really know how to define the wave function renormalization without the full two-point function solution. One may, however, notice two properties if we choose to use the zero-momentum limit at integrand level as in the naive effective potential analysis above. First, the four field vertex at zero-momentum limit, i.e., the second term of the sum in the first line for \eqref{effectivepotential}, remains finite when $D\to 4$. Also, the UV divergence in \eqref{I24}, when compared with usual mass renormalization, receives its corresponding mass dimension via $\beta^{-1}$ instead of $m^2$. One may wander how such modification would affect the usual renormalization group analysis. All these observations seem to hint that Snyder deformed $\phi^4$ theory could possess more complicated quantum behavior than the UV/IR mixing analyzed in the prior sections and require analysis beyond the fixed order loop calculations too.
%to resolve two (of the three) standard renormalization conditions
%\begin{gather*}
%Z_\phi^{-1}\Gamma^{(2)}(p=0)=m^2,
%\\
%Z_\phi^{-2}\Gamma^{(4)}(p_i=0)=g.
%\end{gather*}

\section{Conclusions}

In this article we have studied the effect of Hermitian realization of the nonassociative and noncommutative Snyder deformation of the scalar $\phi^4$ quantum field theory, by computing tadpole diagram contribution to the one-loop two-point function. We have shown that the nonassociativity increases the number of different possible terms in the action with respect to the associative case and affects the results at the quantum level.

We have calculated the momentum-conserving tadpole integrals for the three different inequivalent terms $S_1$, $S_2$ and $S_3$ that can appear in the general $\beta$-exact Snyder $\phi^4$ interaction (\ref{S}). They are found to possess remarkably different properties. Of the three integrals $\mathcal I_1$, $\mathcal I_2$ and $\mathcal I_3$ coming from the three terms $S_1$, $S_2$ and $S_3$, $\mathcal I_1$ is finite in all dimensions, $\mathcal I_2$ is finite when $D<4$ but logarithmically divergent when $D=4$, whereas $\mathcal I_3$ is finite for finite momentum, yet exhibits a logarithmic IR divergence when $p\to 0$ in the $D=4$ case.

The integrals $\mathcal I_1,\mathcal I_2$ and $\mathcal I_3$ exhibit uniform divergent behavior in the commutative limit when $\beta\to 0$, as expected. On the other hand, their infrared limits can be quite different: $\mathcal I_1$ and $\mathcal I_2$ are independent of external momentum, and therefore remain unchanged in the IR limit. However, $\mathcal I_3$ is momentum dependent and exhibits a logarithmic infrared divergence when $D=4$. The logarithmic IR
divergence of $\mathcal I_3$ matches the logarithmic UV divergence of $\mathcal I_2$ at $D\to 4$, which, at the integrand level, is exactly the $p\to 0$ limit of $\mathcal I_3$. For this reason we conclude that a new type of UV/IR mixing, induced by nonassociativity on top of noncommutativity, occurs in $\mathcal I_3$ and represents a general quantum feature of Snyder deformed scalar $\phi^4$ field theory at the level of the one-loop two-point function. We also extend our analysis on UV/IR mixing into the momentum nonconserving integrals (\ref{formal}) by obtaining through UV power counting as qualitative evidence that UV/IR mixing also emerge in momentum nonconserving integrals and therefore should be considered as a general property of the Snyder $\phi^4$ quantum field theory.

At present, the problem of computing complete momentum non-conserving parts of the two-point functions is still open. The method presented in Sec III.B. does allow us to integrate over deformed $\delta$ functions explicitly or implicitly, as well as to analyze certain properties of the loop integrand by using the UV power counting. Yet any analytical solution in closed form of the final integration over the loop momentum is still too far to reach. It is without a doubt that knowing some of such solutions would efficiently improve our overall understanding of the one-loop quantum properties of Snyder $\phi^4$ theory. However many questions, both practical ones (like what would be the total sum of UV/IR mixing terms, and/or whether certain cancellation mechanism for UV/IR mixing could emerge), and conceptual ones (like whether the full one-loop corrected two-point function can bear sound meaning as a QFT), unfortunately still remain unanswered. Anyway,  we hope that some of the above issues could be settled in the future.

While the full one-loop two-point function is not yet available, its zero-momentum limit at integrand level can be defined completely. We exploited this fact discussing an analogue of the Coleman-Weinberg effective potential, noticing that finite one-loop results can be obtained analytically for the three-dimensional theory. Also the UV divergence is clearly reduced when $D=4$. The loop correction tends to diverge when $\beta\to 0$ and therefore could become large when $\beta$ is small. We consider these findings as suggestions towards nonperturbative studies on the Snyder $\phi^4$ theory.

\section{Acknowledgments}
This work is supported by the Croatian Science Foundation (HRZZ) under Contract No. IP-2014-09-9582. We acknowledge the support of the COST Action MP1405  (QSPACE). S.~M. and J. Y. acknowledges support by the H2020 Twining project No. 692194, RBI-T-WINNING. S.~M. acknowledges the support from ESI and COST, from his participation to the workshop "Noncommutative geometry and gravity" and wishes to thank H.~Grosse for a discussion. J.~T. and J.~Y. would like to acknowledge the support of W.~Hollik and the Max-Planck-Institute for Physics, Munich, for hospitality, as well as C.~P.~Martin and P.~Schupp for discussions. Also J.~T. would like to thank  E. Seiler for useful discussions, and V.~G.~Kupriyanov and D.~Lust for useful discussions and pointing new Refs. \cite{Hohm:2017cey,Hohm:2017pnh} to us.

\appendix

\section{THE DETERMINANTS}

We present here an evaluation for the determinants $\det\left(\frac{\partial D_\mu}{\partial k_\nu}\right)$ and $\det\left(\frac{\partial D_\mu}{\partial q_\nu}\right)$. We start with an ansatz for $D_\mu(k,q)$, which contains only scalar and vector objects but not pseudovectors and pseudoscalars, i.e.
\begin{equation}
D_\mu(k,q)=f\left(k^2,k\cdot q, q^2\right)k_\mu+g\left(k^2,k\cdot q,q^2\right)q_\mu.
\label{A.1}
\end{equation}
It is then easy to find that
\begin{gather}
\begin{split}
\frac{\partial D_\mu}{\partial k_\nu}=&\delta^\nu_\mu f\left(k^2,k\cdot q, q^2\right)\\&+k_\mu(2k^\nu f^{(1,0,0)}\left(k^2,k\cdot q, q^2\right)+q^\nu f^{(0,1,0)}\left(k^2,k\cdot q, q^2\right))
\\&+q_\mu(2k^\nu g^{(1,0,0)}\left(k^2,k\cdot q, q^2\right)+q^\nu g^{(0,1,0)}\left(k^2,k\cdot q, q^2\right)),
\label{A.2}
\end{split}
\\
\begin{split}
\frac{\partial D_\mu}{\partial q_\nu}=&\delta^\nu_\mu g\left(k^2,k\cdot q, q^2\right)
\\&+q_\mu(k^{\nu}g^{(0,1,0)}\left(k^2,k\cdot q, q^2\right)+2q^\nu g^{(0,0,1)}\left(k^2,k\cdot q, q^2\right))\\&+k_\mu(k^\nu f^{(0,1,0)}\left(k^2,k\cdot q, q^2\right)+2q^\nu f^{(0,0,1)}\left(k^2,k\cdot q, q^2\right)),
\end{split}
\label{A.3}
\end{gather}
where $f^{(1,0,0)}(x,y,z)=\frac{\partial}{\partial x}f(x,y,z)$, $f^{(0,1,0)}(x,y,z)=\frac{\partial}{\partial y}f(x,y,z)$ and $f^{(0,0,1)}(x,y,z)=\frac{\partial}{\partial z}f(x,y,z)$.

Now, using the Sylvester's determinant identity
\begin{equation}
\det (I_m+AB)=\det (I_n+BA),
\label{A.4}
\end{equation}
one can show that
\begin{equation}
\det\left(\delta_\mu^\nu+a_{1_\mu} b_1^\nu+a_{2_\mu} b_2^\nu\right)=1+a_1\cdot b_1+a_2\cdot
b_2+(a_1\cdot b_1)(a_2\cdot b_2)-(a_1\cdot b_2)(a_2\cdot b_1).
\label{A.5}
\end{equation}
Therefore
\begin{gather}
\begin{split}
\det\left(\frac{\partial D_\mu}{\partial k_\nu}\right)=&f^D\Big(1+f^{-1}\left(2k^2f^{(1,0,0)}+k\cdot q(f^{(0,1,0)}+2g^{(1,0,0)})+q^2g^{(0,1,0)}\right)
\\&+2f^{-2}\left(k^2q^2-(k\cdot q)^2\right)\left(f^{(1,0,0)}g^{(0,1,0)}-g^{(1,0,0)}f^{(0,1,0)}\right)\Big),
\end{split}
\label{A.6}
\\
\begin{split}
\det\left(\frac{\partial D_\mu}{\partial q_\nu}\right)=&g^D\Big(1+g^{-1}\left(k^2f^{(0,1,0)}+k\cdot q(g^{(0,1,0)}+2f^{(0,0,1)})+2q^2g^{(0,0,1)}\right)
\\&+2g^{-2}\left(k^2q^2-(k\cdot q)^2\right)\left(g^{(0,0,1)}f^{(0,1,0)}-f^{(0,0,1)}g^{(0,1,0)}\right)\Big).
\end{split}
\label{A.7}
\end{gather}
Finally we insert the Snyder realization~(\ref{snyderrealization}), and after some algebra we get
\begin{equation}
\det\left(\frac{\partial D_\mu}{\partial k_\nu}\right)=\frac{1+\beta q^2}{(1-\beta k\cdot q)^{D+1}}\left(1-\frac{\beta k\cdot q}{1+\sqrt{1+\beta k^2}}\right)^{D-1},
\label{A.8}
\end{equation}
and
\begin{equation}
\det\left(\frac{\partial D_\mu}{\partial q_\nu}\right)=\frac{(1+\beta k^2)^{\frac{D+1}{2}}}{(1-\beta k\cdot q)^{D+1}}.
\label{A.9}
\end{equation}
We list few special values of these two determinants which are relevant for the calculation in the main text:
\begin{gather}
\begin{split}
&k=0{\;\;\:}:\;\det\left(\frac{\partial D_\mu}{\partial k_\nu}\right)=1+\beta q^2,\;\det\left(\frac{\partial D_\mu}{\partial q_\nu}\right)=1,
\\&
q=0{\;\;\:}:\;\det\left(\frac{\partial D_\mu}{\partial k_\nu}\right)=1,\;\det\left(\frac{\partial D_\mu}{\partial q_\nu}\right)=(1+\beta k^2)^{\frac{D+1}{2}},
\\&
k=-q:\;\det\left(\frac{\partial D_\mu}{\partial k_\nu}\right)=\det\left(\frac{\partial D_\mu}{\partial q_\nu}\right)=(1+\beta k^2)^{-\frac{D+1}{2}}.
\end{split}
\label{A.10}
\end{gather}

\section{DIMENSIONAL REGULARIZATION OF $\mathcal I_3$}

In this section we present the detailed evaluation of \eqref{I3D}. We start by rescaling the mass and momenta with respect to $\beta$:
\begin{equation}
L=\ell\sqrt\beta,\;P=p\sqrt\beta,\;M=m\sqrt\beta.
\end{equation}
Then, after a further redefinition $\mathcal{P}=\frac{P}{1+\sqrt{1+P^2}}$, \eqref{I3D} reduces to
\begin{equation}
\begin{split}
\mathcal I_3=&\beta^{1-\frac{D}{2}}\int\frac{d^D L}{(2\pi)^D}\frac{1}{(L^2+M^2)(1+L^2)}\left(1-\mathcal{P}\cdot L\right)^{1-D}.
\end{split}
\end{equation}
The integrand can then be parametrized by two parameters, one for the $L$-quadratic factors while the other for the $L$-linear factor, i.e.
\begin{equation}
\begin{split}
\mathcal I_3=&\beta^{1-\frac{D}{2}}\int\limits_0^1 dx\,\int\limits_0^\infty dy\,\frac{\Gamma\left(D+1\right)}{\Gamma\left(D-1\right)}\int\frac{d^D L}{(2\pi)^D}\frac{y^{D-2}}{\left(L^2+1+x(M^2-1)+y-y\mathcal{P}\cdot L\right)^{D+1}}.
\end{split}
\end{equation}
It is then straightforward to integrate over $L$ using the (Schwinger) $\alpha$ parametrization
\begin{equation}
\begin{split}
\mathcal I_3=&\frac{\beta^{1-\frac{D}{2}}}{\Gamma\left(D-1\right)}\int\limits_0^1 dx\,\int\limits_0^\infty dy\,\int\limits_0^\infty d\alpha \int\frac{d^D L}{(2\pi)^D}y^{D-2}\,\alpha^D
\\&\exp\left[-\alpha\left(\left(L-\frac{y}{2}\mathcal P\right)^2+x(M^2-1)+1+y-y^2\frac{\mathcal{P}^2}{4}\right)\right]
\\=&\frac{\beta(4\pi\beta)^{-\frac{D}{2}}}{\Gamma\left(D-1\right)}\int\limits_0^1 dx\,\int\limits_0^\infty dy\,\int\limits_0^\infty d\alpha\, y^{D-2}\alpha^{\frac{D}{2}}
\exp\left[-\alpha\left(x(M^2-1)+1+y-y^2\frac{\mathcal{P}^2}{4}\right)\right]
\\=&\beta(4\pi\beta)^{-\frac{D}{2}}\frac{\Gamma\left(\frac{D}{2}+1\right)}{\Gamma\left(D-1\right)}
\int\limits_0^1 dx\,\int\limits_0^\infty dy\,y^{D-2}\left(x(M^2-1)+1+y-y^2\frac{\mathcal{P}^2}{4}\right)^{-1-\frac{D}{2}}
\\=&\beta(4\pi\beta)^{-\frac{D}{2}}\frac{\Gamma\left(\frac{D}{2}\right)}{\Gamma\left(D-1)\right)}
\frac{1}{1-M^2}\left(\mathcal F\Big(M^2,-\frac{\mathcal P^2}{4}\Big)-\mathcal F\Big(1,-\frac{\mathcal P^2}{4}\Big)\right)
\end{split}
\end{equation}
where
\begin{equation}
\mathcal F\left(a,b\right)=\int\limits_0^\infty dy\,\frac{y^{D-2}}{\left(a+y+by^2\right)^\frac{D}{2}}=\int\limits_0^\infty dt\,\frac{1}{\left(a t^2+t+b\right)^\frac{D}{2}}.
\label{mathcalF}
\end{equation}
The integral \eqref{mathcalF} can then be expressed in terms of the Gauss hypergeometric function
\begin{equation}
\mathcal F\left(a,b\right)=(4a)^{\frac{D}{2}-1}\frac{2}{D-1}\,_2F_1\left(\frac{D-1}{2},\frac{D}{2};\frac{D+1}{2};1-4ab\right).
\end{equation}
The expansion of $\mathcal F\left(a,b\right)$ within the small $b$ regime can be done by using the analytical continuation formula of hypergeometric functions, which yields a finite expansion in the $D\to 4$ limit,
\begin{equation}
\mathcal F\left(a,b\right)\underset{D\to 4}{\longrightarrow}\frac{1}{b}+4a\sum\limits_{n=0}^\infty\,(4ab)^n\frac{\Gamma\left(n+\frac{3}{2}\right)}{\Gamma\left(\frac{1}{2}\right)\Gamma\left(n+1\right)}\left(\psi\Big(n+\frac{3}{2}\Big)
-\psi\left(n+1\right)+\ln(4ab)\right).
\end{equation}
Thus the IR limit of $\mathcal I_3$ at $D=4$ boils down to
\begin{equation}
\mathcal I_3=-\frac{1}{16\pi^2\beta}\ln (-\beta p^2)+\mathcal O(1).
\end{equation}
By using the same method it is straightforward to show that,  when  $D=3$,  in the zero external momentum limit $\mathcal I_3$ does converge to $\mathcal I_2$ from \eqref{I23}.

\end{document}